\documentclass{ws-procs975x65}

\begin{document}

\title{NUCLEOSYNTHESIS IN THE GAMMA-RAY BURST ACCRETION DISKS AND ASSOCIATED OUTFLOWS}

\author{INDRANI BANERJEE,~~BANIBRATA MUKHOPADHYAY}

\address{Department of Physics, Indian Institute of Science,\\
Bangalore 560012\\
E-mail: indrani@physics.iisc.ernet.in , bm@physics.iisc.ernet.in\\
}

\begin{abstract}
We investigate nucleosynthesis inside the gamma-ray burst (GRB) accretion disks formed by the Type II 
collapsars and outflows launched from these disks. We deal with accretion 
disks having relatively low accretion rates: $0.001 M_{\odot} s^{-1} \lesssim \dot{M} \lesssim 0.01 M_{\odot} 
s^{-1}$ and hence they are predominantly advection dominated.
We report the synthesis of several unusual nuclei like $^{31}$P, $^{39}$K, $^{43}$Sc, $^{35}$Cl 
and various 
isotopes of titanium, vanadium, chromium, manganese and copper in the disk.
We also confirm that isotopes of iron, cobalt, 
nickel, argon, calcium, sulphur and silicon get synthesized in the disk, as shown by previous authors. 
Much of these 
heavy elements thus synthesized are ejected from the disk and survive in the outflows. 
Indeed, emission lines of many of these heavy 
elements have been observed in the X-ray afterglows of several GRBs.
\end{abstract}


\bodymatter

\section{Introduction}\label{aba:sec1} 
The collapsar model is the most promising theoretical model explaining the long-duration gamma-ray bursts 
(GRBs)
and the supernova (SN)$-$gamma-ray burst connections \cite{MWH01}. We plan to investigate the nucleosynthesis in 
the accretion disks formed by the Type II collapsars\cite{Matsuba} where the accretion rate ($\dot{M}$) is: 
$0.001 M_{\odot} s^{-1} \lesssim \dot{M} \lesssim 0.01 M_{\odot} 
s^{-1}$, when $M_{\odot}$ indicates solar mass, as this regime is the site for the synthesis 
of heavy elements. 
Neutrino cooling becomes important in the inner disk where the temperature and density are higher.
We report the synthesis of various 
 elements like $^{31}$P, $^{39}$K, $^{43}$Sc, $^{35}$Cl and several  
uncommon isotopes of titanium, vanadium, chromium, manganese and copper in the disk for the first time,
to the best of our knowledge, 
apart from isotopes of iron, 
cobalt, nickel, calcium, argon, sulphur and silicon which have already been reported earlier 
\cite{Matsuba}.

We also consider nucleosynthesis in the outflows from these disks using an adiabatic one-dimensional 
outflow model \cite{Fujimoto} and report that many of the heavy elements thus synthesized in the disk do survive in the 
outflow. Additionally, depending on whether significant $^{56}$Ni is synthesized in the outflow, it can be 
predicted whether the outflow will lead to a supernova explosion or not. 

\section{Disk and Outflow Models}
We adopt height-averaged equations based on a 
pseudo-Newtonian 
framework \cite{Mukhopadhyay}. The accretion disk formed in a Type II collapsar is modelled within the 
framework suggested by previous 
authors, \cite{Kohri} where the electron degeneracy pressure and the evolving neutron to proton 
ratio 
are appropriately calculated.  

To investigate nucleosynthesis in the outflow, we adopt a simple, one-dimensional, spherically
expanding and adiabatic outflow model \cite{Fujimoto}.

\section{Nucleosynthesis inside accretion disks and outflows}
We use He-rich and Si-rich abundances as the initial conditions of nucleosynthesis at the outer disk.
Well tested nuclear network code, 
\cite{MC2000} has been used.
 We have modified this code  
further by increasing the nuclear network and including reaction rates from the JINA Reaclib Database,
https://groups.nscl.msu.edu/jina/reaclib/db/ \cite{Cybert}.

We also consider outflow from various radii of ejection, $R_{ej}$, with $R_{ej} < 200 R_g $, and evaluate
the abundance evolution in the outflow assuming the initial composition the same as in the accretion 
disk at $R_{ej}$. 

Figure 1(a) models the abundance evolution around a $3M_\odot$ Schwarzschild black hole accreting at  
$\dot{M}=0.001M_\odot s^{-1}$, with the viscosity parameter $\alpha=0.01$ and the composition of the accreting gas 
at the outer disk similar to the pre-SN He-rich layer. It depicts that the disk  
comprises of several zones characterized by the dominant elements. 
The region $\sim 1000-300 R_{g}$, $R_g$ being Schwarzschild radius, is mainly the $^{40}$Ca, 
$^{44}$Ti and $^{48}$Cr rich 
zone. This is because unburnt $^{36}$Ar undergoes 
$\alpha$-capture reaction to give rise to $^{40}$Ca through $^{36}\rm Ar(\alpha,\gamma)^{40}Ca$ which 
further undergoes $\alpha$-capture reactions to give rise to $^{44}$Ti and $^{48}$Cr.

Inside this region, the temperature and density in the disk favor complete 
photodisintegration 
of $^{44}$Ti and $^{48}$Cr releasing $\alpha$-particles and resulting in the formation of $^{40}$Ca,  
$^{36}$Ar, $^{32}$S and $^{28}$Si, as is evident from Fig. 1(a).
Subsequently, $^{28}$Si and $^{32}$S start burning, which favors formation of 
iron-group elements via {\it photodisintegration rearrangement} reactions \cite{Clayton}.
Therefore, in the range $\sim 300-80 R_{g}$, there is a zone overabundant 
in $^{56}$Ni, $^{54}$Fe, $^{32}$S and $^{28}$Si. 
Inside this zone, all the heavy elements photodisintegrate to $^{4}$He, neutron and proton.

On taking the same disk and black hole parameters but considering Si-rich abundance at the outer disk we 
note 
that the disk has a huge zone rich in $^{28}$Si and $^{32}$S extending upto radius $\sim 250R_g$. 
Inside this radius, silicon burning commences and soon the disk becomes rich in $^{54}$Fe, $^{56}$Ni and
$^{58}$Ni. Inside $\sim 70 R_{g}$, all the
heavy elements again get photodisintegrated to $^{4}$He, protons and neutrons.
Another remarkable feature in the He-rich and Si-rich disks is that inside $\sim 100 R_{g}$, the 
abundances of various elements start becoming almost identical as if once threshold density and 
temperature are achieved, the nuclear reactions forget their initial abundances and follow only the 
underlying disk hydrodynamics. 

On increasing $\dot{M}$ ten times we find that the individual zones
shift outward retaining similar composition as is in the low $\dot{M}$ cases described above.

Next we consider the outflow from $40 R_g$ of the He-rich disk with $\dot{M}=0.001M_{\odot}s^{-1}$, as 
described above. We show 
that $^{56}$Ni is abundantly synthesized in the outflow along with isotopes of copper and zinc, 
as shown in Fig. 1(b). Synthesis of $^{56}$Ni  
signifies that the outflow will lead to a 
supernova 
explosion. Figure 1(b) also depicts that on changing the initial velocity of ejection the
final abundances of the nucleosynthesis products change significantly. More $^{56}$Ni is 
synthesized when the velocity of ejection is low (see Fig. 1(b)) because then the temperature drops slowly 
in the ejecta which facilitates greater recombination of alphas to nickel.  
 
\begin{figure}[t]
\vskip-5cm
\begin{center}
\psfig{file=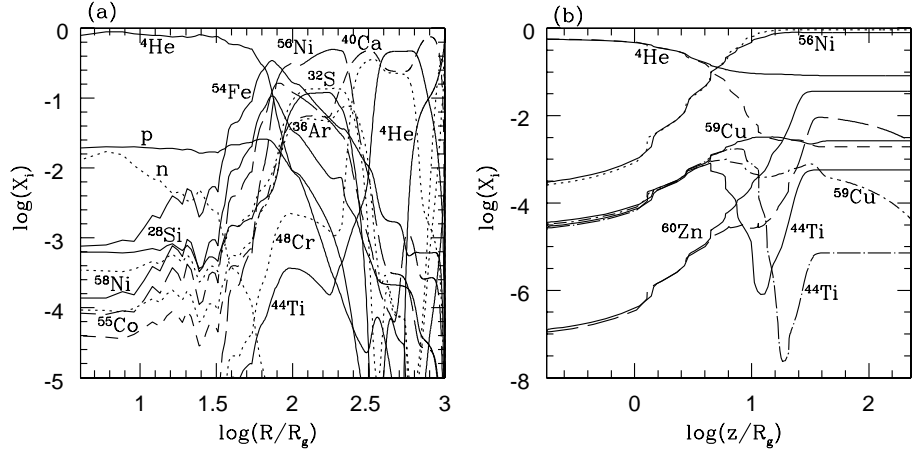,width=5in}
\end{center}
\vskip-2.2cm
\caption{(a) Zones characterized by dominant elements in the  
disk
with He-rich abundance at the outer disk. (b) Abundance 
evolution in the outflow from $R_{ej} \sim 40 R_g$ of the disk in (a),
solid lines correspond to the higher velocity of ejection and other lines to
lower velocity, in each set.}
\label{aba:fig1}
\end{figure}

\section{Conclusions} 
We report for the first time, to the best of our knowledge, that several unusual nuclei like $^{31}$P, $^{39}$K, $^{43}$Sc, $^{35}$Cl 
and various uncommon
isotopes of titanium, vanadium, chromium, manganese and copper get synthesized in the disk apart from 
isotopes of iron, cobalt, nickel, silicon, sulphur, argon and calcium. 
Several zones, characterized by dominant elements, are formed in the disk. 
The elemental distribution in the disk
starts appearing identical once threshold density and temperature are reached. The respective zones shift 
outwards on increasing the mass accretion rate. 
Several of these heavy elements survive in the outflow from these disks, and when $^{56}$Ni is
abundantly synthesized in the outflow, there is always a supernova explosion. \\   

 This work was partly supported by the ISRO grant ISRO/RES/2/367/10-11.

\end{document}